\documentclass[showpacs,preprintnumbers,amsmath,amssymb]{revtex4}
\usepackage{graphicx}
\usepackage{dcolumn}
\makeatletter
\input{epsf}

\newcommand{\bd}{\begin{document}}
\newcommand{\ed}{\end{document}}
\newcommand{\bc}{\begin{center}}
\newcommand{\ec}{\end{center}}
\newcommand{\bfr}{\begin{flushright}}
\newcommand{\efr}{\end{flushright}}
\newcommand{\vs}{\vspace}
\newcommand{\hs}{\hspace}
\newcommand{\beq}{\begin{equation}}
\newcommand{\eeq}{\end{equation}}
\newcommand{\lb}{\linebreak}
\newcommand{\mb}{\makebox}
\newcommand{\fb}{\framebox}
\newcommand{\mc}{\multicolumn}
\newcommand{\un}{\underline}
\newcommand{\lefq}{\lefteqn}
\newcommand{\ba}{\begin{array}}
\newcommand{\ea}{\end{array}}
\newcommand{\beqa}{\begin{eqnarray}}
\newcommand{\eeqa}{\end{eqnarray}}
\newcommand{\beqas}{\begin{eqnarray*}}
\newcommand{\eeqas}{\end{eqnarray*}}
\newcommand{\bfg}{\begin{figure}}
\newcommand{\efg}{\end{figure}}
\newcommand{\bds}{\begin{displaymath}}
\newcommand{\eds}{\end{displaymath}}
\newcommand{\btb}{\begin{tabbing}}
\newcommand{\etb}{\end{tabbing}}
\newcommand{\para}{\parallel}
\newcommand{\pad}{\partial}
\newcommand{\nn}{\nonumber}
\newcommand{\la}{\leftarrow}
\newcommand{\ra}{\rightarrow}
\newcommand{\lgla}{\longleftarrow}
\newcommand{\lgra}{\longrightarrow}
\newcommand{\La}{\Leftarrow}
\newcommand{\Ra}{\Rightarrow}
\newcommand{\Lra}{\Leftrightarrow}
\newcommand{\Lgla}{\Longleftarrow}
\newcommand{\Lgra}{\Longrightarrow}
\newcommand{\bm}{\boldmath}
\newcommand{\lan}{\langle}
\newcommand{\ran}{\rangle}
\renewcommand{\a}{\alpha}
\renewcommand{\b}{\beta}
\newcommand{\g}{\gamma}
\newcommand{\G}{\Gamma}
\renewcommand{\d}{\delta}
\newcommand{\eps}{\epsilon}
\newcommand{\s}{\sigma}
\newcommand{\lam}{\lambda}
\newcommand{\D}{\Delta}
\newcommand{\vare}{\varepsilon}
\newcommand{\pr}{\prime}
\newcommand{\ro}{\rho}
\newcommand{\nab}{\nabla}
\newcommand{\m}{\mu}
\newcommand{\n}{\nu}
\newcommand{\Sg}{\Sigma}
\newcommand{\p}{\pi}
\newcommand{\R}{I\!\!R}
\newcommand{\om}{\omega}
\newcommand{\Om}{\Omega}
\newcommand{\ze}{\zeta}
\newcommand{\vart}{\vartheta}
\newcommand{\tri}{\triangle}
\newcommand{\f}{\frac}
\newcommand{\iny}{\infty}
\newcommand{\pro}{\propto}
\begin{document}
\title{The Berry phase in ferromagnetic spin systems and anomalous Hall Effect}

\author{B. Basu}
\email{banasri@isical.ac.in}
\author{P. Bandyopadhyay}
 \email{pratul@isical.ac.in}
\affiliation{Physics and Applied Mathematics Unit\\
 Indian Statistical Institute\\
 Kolkata-700108 }

\begin{abstract}
We have shown that the study of topological aspects of the 
underlying geometry in a ferromagnetic spin system gives rise to an
intrinsic Berry phase. This real space Berry phase arises due to the spin rotations of conducting electrons which can be manifested as a further contribution in anomalous Hall effect.

\end{abstract}

\maketitle
 In ferromagnetic metals with a spontaneous broken time reversal symmetry, 
 besides the ordinary Hall effect linear dependence of the off diagonal resistivity on the appilied magnetic field, there exists a contribution proportional to the magnetization of the ferromagnet.
The transverse resistivity $\rho_H$ in ferromagnets consists of the
two contributions, \beq \rho_H~=~ R_0B+ R_SM \eeq where $B$, $M$,
$R_0$ and $R_S$ are magnetic induction, magnetization, ordinary Hall
coefficient and anomalous Hall coefficient. The second term, which
is proportional to the magnetization, represents the anomalous Hall
effect (AHE). Conventionally, the AHE is ascribed to spin-orbit
interaction which involves a coupling of orbital motion of electrons to the spin polarization of conduction electrons \cite{so}
or to the asymmetric skew scattering of conduction electrons by the
fluctuation of localised moments \cite{skew}. 

In a recent paper Haldane, \cite{h}
has pointed out that the intrinsic AHE in metallic ferromagnets is
controlled by Berry phases accumulated by adiabatic motion of
quasiparticles on the Fermi surface in the presence of broken
inversion or time -reversal symmetry. 
 This theory was in support of a report of
Fang $et.$ $al$\cite{f} where they have shown that the AHE is
related to the Berry phase acquired by the Bloch wave functions when
the magnetic "monopole" is in the crystal momentum space. The
topological features associated with the AHE have also been studied
by other authors \cite{y,t}. 
It has been realized that under appropriate conditions a chirality contribution shows up in AHE \cite{y,t}. In case of canonical spin glass chirality contribution to the AHE was examined by Tatara and Kawamura \cite{t,k}. 
 However, in order to get a
net topological field (chirality) the spin-orbit coupling must be
invoked. Besides, we have problems related to the $2$D Kagome
lattice or $3$D Pyrochlore lattice where the net topological field
vanishes though a nonvanishing topological Hall effect may be
obtained \cite{t,6}. Breno, Dugaev and Taillefumier \cite{7}have
shown that a net topological field can be obtained by means of some
external parameter. The analysis of this topological Hall effect
does not depend on spin-orbit coupling but arises solely from the
Berry phase acquired by an electron moving in a smoothly varying
magnetization.

Recently, the AHE has been studied in a series of AuFe samples
\cite{cam}. It has been observed that below a critical Fe
concentration the alloys are spin glasses while for higher
concentrations the alloys have been dubbed `re-entrant'; on cooling
one first encounters a ferromagnetic ordering temperature $T_c$ and
then a second canting temperature $T_k$ which is signalled by a
dramatic drop in the low field ac susceptibility. Below $T_k$ the
system still has an overall ferromagnetic magnetization but the
individual Fe spins become canted locally with respect to the global
magnetization axis. The experimental data demonstrate that the
degree of canting strongly modifies the AHE. This is a physically
distinct Berry phase contribution occurring in real space when the
spin configuration is topologically nontrivial. The present paper is an attempt to address  some aspects associated with the geometry of
this type of ferromagnetic system  and show that the contribution of
AHE in this system is a topological effect which arises due to the Berry curvature accumulated by spin rotations of moving electrons.

We start with a model which represents the effective ferromagnetic interaction between electrons on different sites. The Hamiltonian is given by
\beq
H=\sum_{i,j}tc_i^\dagger c_j~-~\frac{J_H}{2}\sum_i{\bf S}_i.[{c_i}^\dagger \vec{\sigma} c_j]
\eeq
where $(i,j)$ runs the nearest neighbour sites, $c_i(c_i^\dagger)$ is the annihilation (creation) operator at the site $i$ and ${\bf S}_i$ is the classical spin localised at the site $i$ and $\vec{\sigma}$ are Pauli matrices. This describes an electron hopping from site $i$ to site $j$ coupled to a spin at each site with a Hund coupling $J_H$. When $J_H$ is strong enough the spin of the hopping electron is forced to align parallel to ${\bf S_i}$ and ${\bf S_j}$ at each site.  

For a ferromagnetic system which allows a ferromagnetic ordering
below $T_c$ as well as canting of local spins below $T_k$
we shall focus only on the position dependent magnetization direction 
\beq {\bf n}=\frac{{\bf M}({\bf r})}{M}
\eeq 
with a fixed magnitude $|\bf{M}|$=$M$.
The magnetization direction  ${\bf n}({\bf r}, t)$ satisfies the equation of motion
\beq
\partial_t {\bf n}({\bf r}, t)=-\gamma {\bf n}({\bf r}, t) \times H_{eff}(\bf{r})
\end{equation}
where $\gamma$ is (minus) the gyromagnetic ratio $H_{eff}(\bf{r})$ is the so-called $effective$ $magnetic$ $field$ depending on the magnetization distribution $\bf{M(r)}$.
Indeed, we can use a gauge transformation
$U({\bf r})$ which makes the quantization axis oriented along the
vector ${\bf n}({\bf r})$ at each point so that we can write \beq
U^{\dagger}({\bf r})[\vec{\sigma}.{\bf n}({\bf r})]U({\bf
r})=\sigma_z \eeq
In generalized spin
systems we may assume that the direction of the vector ${\bf n}({\bf
r})$ is arbitrary at different sites. To view the rotation of
the magnetization vector such that the vector ${\bf n}({\bf r})$ is
oriented along the quantization axis we may consider the vector
${\bf n}({\bf r})$  is rotating with an angular velocity $\omega_0$
around the $z$-axis under an angle $\nu$ so that at any instant of
time, the magnetization vector is at a position which makes an
arbitrary angle $\nu$ with the quantization axis. Then we
can write the unit vector in terms of the variables $\nu$ and $t$
as \beq {\bf n}({\bf r})\rightarrow {\bf n}(\nu,t) \eeq where  
\begin{equation}\label{nn}
    {\bf n}(\nu,t)=
    \left(
    \begin{array}{cc}
    \sin \nu & \cos(\omega_0 t) \\
    \sin \nu & \sin(\omega_0 t) \\
    \cos \nu
    \end{array}\right)
\end{equation}
It may be added here that there may be an additional phase factor in a multi-spin system where $\cos(\omega_0t)$should be replaced by $\cos(\omega_0t+\phi)$. However as this will not change the physics we are considering, we may omit it here.
The instantaneous eigenstates of a spin operator in direction
 ${\bf n}(\nu,t)$ expanded in the $\sigma_z$-basis are given by
\begin{equation}\label{art}
\begin{array}{ccc}
    \displaystyle{|\chi(\uparrow))_n;t>}&=&\displaystyle{\cos \frac{\nu}{2} |\uparrow_z> +~ \sin
    \frac{\nu}{2} e^{i\omega_0 t}|\downarrow_z> }\\
    &&\\
    \displaystyle{|\chi(\downarrow))_n;t>}&=&\displaystyle{-\sin \frac{\nu}{2} |\uparrow_z> +~ \cos
    \frac{\nu}{2} e^{i\omega_0 t}|\downarrow_z>}
\end{array}
\end{equation}

For the time  evolution from $t=0$ to $t=\tau$ where
$\tau=\displaystyle{\frac{2\pi}{\omega_0}}$ each eigenstate will
pick up a geometric phase (Berry phase) apart from the dynamical
phase which is of the form 
\begin{eqnarray}\label{upn}
\displaystyle{|\chi(\uparrow)_n;t=0>\rightarrow|\chi(\uparrow)_n;t=\tau>=e^{i\gamma_+
(\nu)}
~e^{i\theta_+} |\chi(\uparrow)_n;t=0>} \nonumber \\
\nonumber \\
\displaystyle{|\chi(\downarrow)_n;t=0>\rightarrow|\chi(\downarrow)_n;t=\tau>=e^{i\gamma_-
(\nu)} ~e^{i\theta_-} |\chi(\downarrow)_n;t=0>}
\end{eqnarray}
where $\gamma_\pm$ is the Berry phase which is half of the solid
angle $\frac{1}{2}\Omega$ swept out by the magnetization vector and
$\theta_\pm$ is the dynamical phase. The dynamical phase can be eliminated by the spin-echo method \cite{beren}. However, in the context of AHE we are only concerened with the Berry phase. Geometrically we may
get the explicit values of the Berry phase given by $\gamma_\pm$ as
\begin{equation}\label{gpn}
    \begin{array}{cl}
      \displaystyle{\gamma_+(\nu)}~= & \displaystyle{-\pi(1-\cos \nu)} \\
      &\\
      \displaystyle{\gamma_-(\nu)}~= & \displaystyle{-\pi(1+\cos \nu)=-\gamma_+(\nu)-2\pi} \\
    \end{array}
    \end{equation}
This equation helps us to note that in any arbitrary spin
system the Berry phase aquired by a spin up eigenstate in direction ${\bf n}(\nu)$ is given by
\beq \gamma_+(\nu)~= -\pi(1-\cos \nu) \eeq

 It is known that a free polarized spin can be represented by a two-component spinor which is achieved when a scalar particle is attached with one magnetic flux
quantum. Now in unit of magnetic flux quantum  
$\frac{hc}{|e|}$ the corresponding magnetic (monopole) strength is given
by $\mid\mu\mid=\frac{1}{2}$. So the phase acquired by a free
polarized spin encircling a closed path may be represented by the
phase acquired by a scalar particle encircling a magnetic flux
quantum which is given by $e^{i2\pi\mu}$ with $\mu=1/2$. Indeed this
phase $e^{i\pi}$ corresponds to the phase acquired by a fermion
after $2\pi$ rotation.

From eqn.(11) we note that in a multi-spin system,
the Berry phase acquired by a spin-up eigenstate in direction ${\bf
n}(\nu)$ can be written in terms of $\mu$ as
 \beq \label{g} e^{i\gamma_+(\nu)}=e^{i2\pi\m(1-cos\nu)}  \eeq  ~~~with  $\m=1/2$

This helps us to note that in a spin system there is a deviation of the phase factor
acquired by a spin eigenstate in direction ${\bf n}(\n)$from that of
a free polarized spin. The deviation of the phase factor
$\mid\Delta\g\mid$ from the free spin case is given by \beq
\label{dev} \mid\Delta\g\mid=\frac{1}{2}\cos\nu \eeq Now for a
ferromagnetic system where all the spins are aligned along the
quantization axis ($\nu=0$) we have the intrinsic phase factor
acquired by a spin eigenstate \beq \mid\Delta\g\mid=\frac{1}{2} \eeq
This intrinsic phase factor acquired by the spin eigenstate in a ferromagnetic system, is the manifestation of some inherent magnetic field in these types of   
systems. This fictitous magnetic field is present due to the effect of the Berry phase of the localised spins and conduction electrons move in this field 
 when there is strong Hund coupling.

Now from eqn.(\ref{dev})we observe that we will have variation of
$|\Delta \g|$ depending on the angle $\nu$ which is associated with
the degree of canting of the local spins. In fact, the system will
still have an overall ferromagnetic magnetization but the degree of
canting of individual spins will now change the inherent magnetic
field associated with $|\Delta \g|$ inducing a modification of the
AHE as observed in experinents \cite{cam}
We  introduce a gauge potential corresponding to this magnetic field  
\beq {\bf
A}({\bf r},t)=-2\pi i \phi_0~U^\dagger({\bf r},t)\partial_i~U({\bf r},t)
\eeq
 where
$\phi_0=\frac{hc}{|e|}$ is the flux quantum and $i=x,y$ .
Now if we neglect the spin-flip transitions and assume that the system is in  the spin up (or, spin down)subspace only , we can substitute the matrix gauge field by an Abelian gauge field ${\bf a}({\bf r}, t)$. In fact, we
can choose ${\bf a}({\bf r}, t)$ as 
\beq \label{a} {\bf a}({\bf r},t)=i<{\bf
n}({\bf r}, t)\mid\nabla{\bf n}({\bf r}, t)> \eeq 
The Berry phase for a closed path $\Gamma$ is given by
\beq
\Gamma=exp i \oint_\Gamma {\bf a}(\bf{r})d\bf{r}
\eeq
If we assume conduction electrons as to represent $2D$ electron gas we can consider the continuum limit.
Writing \beq
a_i({\bf r}, t)=\frac{n_x
\partial_i n_y-n_y\partial_i n_x}{1+n_z} \eeq  the
corresponding field $B$ takes the form \beq
B=\frac{1}{4\pi}\epsilon_{\mu\nu\lambda}n_\mu(\partial_xn_\n)(\partial_yn_\lambda)
\eeq 
The integral of this topological field over an area enclosed by
an arbitrary contour is proportional to the Berry phase.

The  topological current can be defined as  \beq
J_\m=\frac{1}{8\pi}\epsilon_{\mu\nu\lambda}{\bf n}(\partial_\nu {\bf
n}\times\partial_\lambda{\bf n}) \eeq In terms of the vector
potential $a_\m$ the topological current is given by
\beq
J_\m=\epsilon_{\m\n\lambda}\partial_\n a_\lambda \eeq 
 This is generated due to the Berry phase accumulated by the spin rotations of moving electrons  when the background magnetization is not uniform in space. 



 In terms of the Abelian gauge field, 
 the Hamiltonian for a $2D$ electron gas may be expressed as
 \beq
H=-\frac{\hbar^2}{2m}[\frac{\partial}{\partial {\bf r}}-ie{\bf
a}({\bf r})]^2-gM\sigma_z \eeq
and  we can define the covariant momentum operator
 \beq
 \pi_\m=-i\frac{\partial}{\partial x_\m}-ea_\m({\bf r})
 \eeq
 This leads to the noncomutativity of the momentum components 
 \beq
 [\pi_x,\pi_y]=ie(\partial_x a_y -\partial_y a_x)=ieB_z \eeq
This ($B_z$) is the $inherent$ $magnetic$ $field$ we were talking about. 
We may add here that a new form of non-commuatative space can be formulated where noncommutativity in the momentum space induces a singular type of magnetic field in the real space \cite{subir,sb}. 

To conclude, we may say 
that for a spatially varying magnetization in a ferromagnetic spin system a topological current is generated 
due to the topological properties associated with the underlying geometry of the system. The inherent magnetic type of behavior is caused by the Berry curvature 
 which arises due to the spin rotations of conducting electrons and is the effect of  noncommutativity in momentum space. 
 



 \ed

The relevance of the non-coplanar spin
configurations to the AHE has been discussed in the context of
perovskite type manganites at high temperatures\cite{y}. In metallic
magnets a complex phase factor occurs when an electron hops along
the noncoplanar spin configuration and the effective magnetic field
is represented by the net spin chirality. This $internal$ $magnetic$
$field$ is expected to manifest itself in the AHE \cite{y,t,5}.

Introducing the polar coordinates we can write 
\beq
<\chi_i|{\bf S}_i|\chi_j>=\frac{1}{2}(\sin\theta_i \cos\phi_i,~\sin\theta_i\sin\phi_i,~\cos\theta_i)
\eeq      
where 
\beq
|\chi_i>=[\cos\f{\theta_i}{2},~\sin\f{\theta_i}{2}e^{i\phi_i}]
\eeq
is the spin wave function.  
Let us now think otherway and proceed in a semiclassical framework.